\documentstyle[psfig]{l-aa}

\begin{document}

\thesaurus{13.25.2,11.14.1,11.19.1,03.13.6}

\title  { The Seyfert Galaxy NGC 6814 --- a highly variable X-ray source }
\author {       Michael K\"onig\inst{1} \and
		Susanne Friedrich\inst{1} \and
		R\"udiger Staubert\inst{1} \and
                Jens Timmer\inst{2,3}   }

\institute{
	       Institut f\"ur Astronomie und Astrophysik,
	       Astronomie,
	       Universit\"at T\"ubingen,
	       Waldh\"auser Str. 64,
	       D -- 72076 T\"ubingen
               \and
	       Fakult\"at f\"ur Physik,
	       Albert-Ludwigs-Universit\"at, 
	       Hermann-Herder Str. 3,
	       D -- 79104 Freiburg
               \and
               Freiburger Zentrum f\"ur Datenanalyse und Modellbildung,
               Albert Str. 26-28,
               D -- 79104 Freiburg
	}

%\date { }
%
\maketitle
\begin{abstract}
The Seyfert galaxy NGC~6814 is a highly variable X-ray source despite the
fact that it has recently been shown not to be the source of periodic
variability. The 1.5 year monitoring by ROSAT has revealed a long term
downward trend of the X-ray flux and an episode of high and rapidly varying
flux (e.g. by a factor of about 3 in 8 hours) during the October 1992 PSPC
observation. Temporal analysis of this data using both Fourier and
autoregressive techniques have shown that the variability timescales 
are larger than a few hundred seconds. The behavior at higher
frequencies can be described by white noise.

\end{abstract}
\section {Introduction}

The Seyfert galaxy NGC~6814 was long believed to be the only firm example
of an Active Galactic Nuclei (AGN) with a persistent periodicity in the
X-ray flux of about 3.4 hours (Mittaz and Branduardi 1989, Done et
al. 1990). From ROSAT observations in October 1992 (Staubert et al. 1994)
and April 1993 (Madejski et al. 1993) it was concluded that a so far
unknown galactic object in the near vicinity of the Seyfert galaxy was
emitting the periodically modulated flux. This new object
RX$\,$J1940.1-1025 which is located about 37 arc min to the west of
NGC~6814 was identified as an asynchronous magnetic cataclysmic
variable (Staubert et al. 1994; Patterson et al. 1995; Friedrich et
al. 1996; Watson et al. 1995).

X-ray flux from the direction of NGC~6814 was dicovered by ARIEL V in 1974
(Cooke et al. 1978) and identified with the Seyfert galaxy which was
the most prominent optical counterpart in the satellites field-of-view,
eventhough, the likelihood of this correlation was not very good (Elvis
et al. 1978).

Considering the limited spatial resolution of the collimated proportional
counter instruments ARIEL V, HEAO-A2, Ginga LAC and EXOSAT ME (FOV of
$1\rlap{.}^\circ 9\times 0\rlap{.}^\circ 4$, $1\rlap{.}^\circ 7\times
0\rlap{.}^\circ 3$, $1\rlap{.}^\circ 1\times 2\rlap{.}^\circ 0$,
$0\rlap{.}^\circ 75\times 0\rlap{.}^\circ 75$, FWHM respectively),
the misidentification of the periodic source is not surprising and an
imaging instrument such as ROSAT was needed to solve the puzzle.

ROSAT is able to study the radiation from NGC~6814 as an individual source
without contamination. NGC~6814 is not an X-ray source with constant X-ray
brightness, but shows strong variability. We have analyzed all ROSAT PSPC
observations of the galaxy using Fourier techniques (Scargle 1982) and
Linear State Space Models, denoted as SSM (Koen and Lombard 1993, K\"onig
and Timmer 1996). Although NGC~6814 is no more a periodic source, it
remains one of the most variable extragalactic X-ray sources.

\section {ROSAT Observations}

ROSAT has performed four pointed observations of NGC~6814 in April and
October 1992 and 1993, respectively. The countrate of NGC~6814 varies
between $0.007$ and $0.177$ counts/sec (1$\sigma$ errors) within the ROSAT
PSPC observations (0.1 -- 2.4 keV energy range) and the mean countrate of
$0.092$ counts/sec corresponds to an X-ray flux of $1.35 \cdot
10^{-12}$erg/s/cm$^{2}$ assuming a single power law with fixed $N_{\rm{H}}
= 0.98 \cdot 10^{21}$ cm$^{2}$ (galactic 21cm column density taken from
Elvis et al. 1989).  The mean 1 keV intensity is $5.8 \cdot 10^{-4}$
phot/cm$^{2}$/s/keV with a photon index $\Gamma = 1.47$ in the 0.1 -- 2.4
keV energy range (see Fig.1).

Data reduction and spectral model fitting was done within the MIDAS/EXSAS
package (provided by ESO/MPE Garching). The source counts were extracted
from the raw data within a circle centered on the source (with a radius of
400 -- 600 sky pixels corresponding to source intensity) and the background
counts were derived from a ring around that source circle (all visible
X-ray sources within this background ring have been removed). Deadtime and
effective area correction has been applied to the data set. Subsequently,
the time dependent background was subtracted from the source lightcurve.

\begin{figure}
%\picplace{8 cm}
\centering
\psfig{file=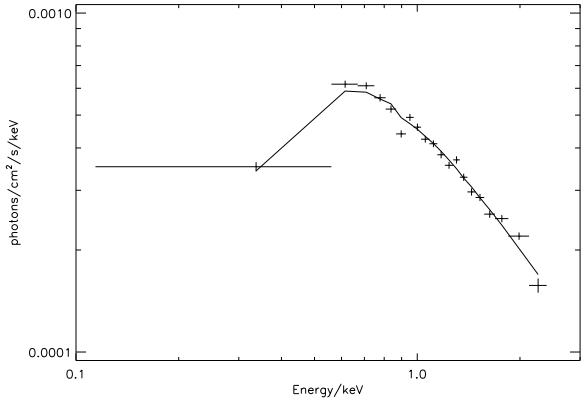,width=8cm}
\caption[ ]{Energy spectrum of merged ROSAT data from NGC~6814 (see Section
2 for the parameters of the power law fit).  }
\end{figure}

\section {Flux and spectral variability}

The long term behavior of the X-ray flux as measured by ROSAT exhibits a
downward trend with a flare-like feature in the October 1992 data
(Fig.2). Over the 1.5 year time baseline of the four observations the
mean countrate decreases by about 60\%.  All four individual observations
were tested for variability. Only the ROSAT observation of October 1992
indicates strong flux modulation on timescales of hours (see Tab.1). The
other three observations can be well modelled with a constant X-ray flux
with white noise stochastic variations.
 
We have analyzed the October 1992 observation in detail (Fig.3). The X-ray
flux of NGC~6814 varies by a factor of 2.8 within a timescale of about 8
hours. The observed lightcurves of single orbits indicate variability on
timescales of a few hundred seconds (see Section 4).

\begin{table*}\centering
\caption{Individual ROSAT observations of NGC~6814 with single power law
spectral fits (fixed N$_{H}^a$). Uncertainties are standard errors as given
by the EXSAS routines.}
\begin{tabular}[c]{ccccccccc}
\hline Observation/PI & ${\rm{HJD}}^{b}$ & length & $\rm{\mbox{mean
countrate}}$ & rms & ${{\chi}^{2}_{\rm red}}^{c}$ & ${\Gamma_{\rm soft}}^d$
& ${{\rm{I}}_{\rm 1keV}}^e$ & $\chi^2 (dof)$ \\ & & (ks) & (cts/s) &
(cts/s)& & & \\ \hline Apr 1992 / Staubert & 244~8742.148 & 8.5&0.137 &
0.042 & 1.44 & 1.65 $\pm$ 0.16 & 0.85 $\pm$ 0.04 & 25.2 (19) \\ Oct 1992 /
Staubert & 244~8915.282 & 28.4&0.181 & 0.100 & 7.07 & 1.43 $\pm$ 0.08 &
1.20 $\pm$ 0.03 & 28.2 (29) \\ Apr 1993 / Madejski& 244~9078.858 &
37.1&0.062 & 0.032 & 1.69 & 1.49 $\pm$ 0.13 & 0.39 $\pm$ 0.02 & 7.5 (16) \\
Oct 1993 / Staubert& 244~9276.694 & 28.0&0.037 & 0.044 & 1.27 & 1.45 $\pm$
0.37 & 0.20 $\pm$ 0.02 & 10.3 (13) \\ \hline \multicolumn{8}{l}{\small $^a$
N$_{H}=0.98 \cdot 10^{21}\rm{cm}^{-2} $ (Elvis et al. 1989)}\\
\multicolumn{8}{l}{\small $^b$ mean time of observation}\\
\multicolumn{8}{l}{\small $^c$ reduced ${\chi^2}$ of a test on constant
flux}\\ \multicolumn{8}{l}{\small$^d$ photon index of the fitted power law
spectrum}\\ \multicolumn{8}{l}{\small$^e$ intensity in
$10^{-3}$phot/cm$^2$/s/keV at 1 keV for z=0}\\

\end{tabular}
\end{table*}

We have also investigated the spectral behavior of NGC 6814 by fitting
single and double power laws to the individual ROSAT PSPC observations. The
single power law fits were performed both with free and fixed hydrogen
column density $N_{\rm{H}}$. For the latter we used $0.98 \cdot 10^{21}
\rm{cm}^{-2}$ derived by Elvis et al. (1989). Also double power law fits
were performed for free and fixed $N_H$ with the hard photon index always
fixed to ${\Gamma} = 1.5$ as derived by Turner et al. (1992) from Ginga
observations\footnote{However, one should keep in mind that the obervation
is contaminated by the close X-ray binary RXJ1940.1-1025, which has a
harder energy spectrum (with a power law slope of about 1) in the ROSAT
energy range compared to the spectrum of the Seyfert galaxy NGC~6814.}. The
best fit, i.e. the lowest reduced $\chi^{2}$ value was always obtained for
a single power law with fixed $N_{\rm{H}}$. The results of these single
power law fits for all individual ROSAT observations are listed in Table 1.
  
To test the hypothesis of a spectral hardening with decreasing flux (Yaqoob
et al. 1989) , we separated the October 1992 observation in two groups,
summing all orbits with high countrates ($>$0.25 counts/sec) and with low
countrates ($<$0.15 counts/sec), respectively. Subsequently single power
law spectra with $N_H$ fixed to $0.98 \cdot 10^{21} \rm{cm}^{-2}$ were
fitted to each data group. A photon index of $1.42 \pm 0.11$ for the `low
flux' data and one of $1.48 \pm 0.13$ for the `high flux' data was
obtained. This is consistent with a constant spectral shape while the X-ray
flux varies by a factor of about three.

\begin{figure}
%\picplace{8 cm}
\centering
\psfig{file=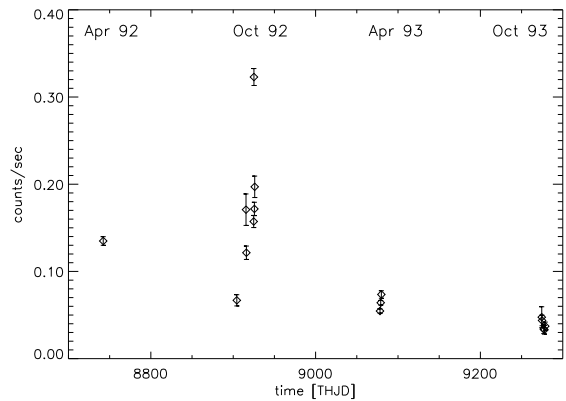,width=8cm}
\caption[ ]{Complete X-ray lightcurve of the Seyfert galaxy NGC~6814 (ROSAT
PSPC, 0.1 -- 2.4 keV energy range). The abscissa denotes the truncated
heliocentric Julian Days. The ordinate values represent mean countrates of
four consecutive orbits.}
\end{figure}

\begin{figure}
%\picplace{8 cm}
\centering
\psfig{file=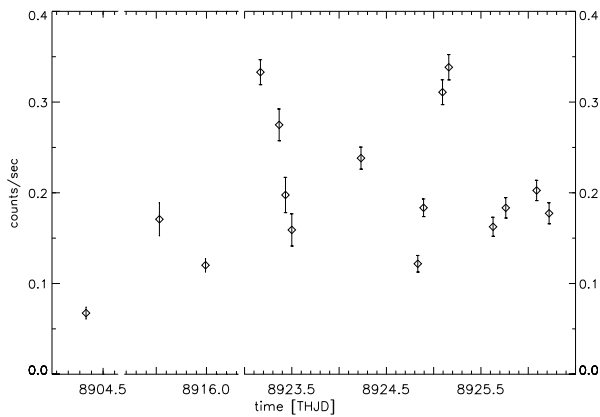,width=8cm}
\caption[ ]{ROSAT PSPC observation of October 1992 (mean countrates for
individual orbits are shown, the abscissa is subdivided into three time
intervalls with 0.1 day grids each).}
\end{figure}

\section {Short term variability}

A common phenomenon of AGN is the strong variability that can be found in
their X-ray lightcurves. This is often described as flickering or
$1/f^{\alpha}$ fluctuations (Lawrence et al. 1987). The $1/f^{\alpha}$ term
describes the distribution of power as a function of frequency in the
periodogram.

Any individual observation of the NGC~6814 can be interpreted as a single
realization or better as an observation of the realization of a stochastic
process. The observed most prominent modulation feature strongly depends on
the observation. Therefore, a charaterization which is based on the
variability on a single timescale (such as the flux doubling time e.g.
McHardy 1988) could be misleading. A more fruitful approach is the
examination of the distribution of all observed timescales which is done by
computing the periodogram.

The lightcurve of the October 1992 ROSAT observation has a very poor duty
cyle of 1.5\%. The duty cycle describes the real on-source time relative to
the temporal baseline of the total observation. The complete observation
consists of 16 orbits covering a time interval of about 21 days. If the
first three orbits are omitted the duty cycle increases to 8.3\% with a
baseline of 260 ks. Due to the large gaps in the ROSAT observation of
NGC~6814, Fourier techniques can only be used with great caution as the
true periodogram might be strongly hampered by the convoluted Fourier
transform of the sampling pattern inducing aliasing and spectral leakage
(Papadakis and Lawrence 1995). In order to estimate this influence we have
simulated 100 white noise lightcurves with the same mean, variance, and
sampling pattern as the observed original ROSAT lightcurve. The
corresponding periodograms have been compiled to a sample periodogram
(Fig.4b) which shows a nearly flat frequency distribution. The influence of
the sampling pattern on the original periodogram can therefore be neglected
and the increase of power towards low frequencies (see Fig.4a) is a real
temporal effect of the X-ray source.

\begin{figure}
%\picplace{8 cm}
\centering
\psfig{file=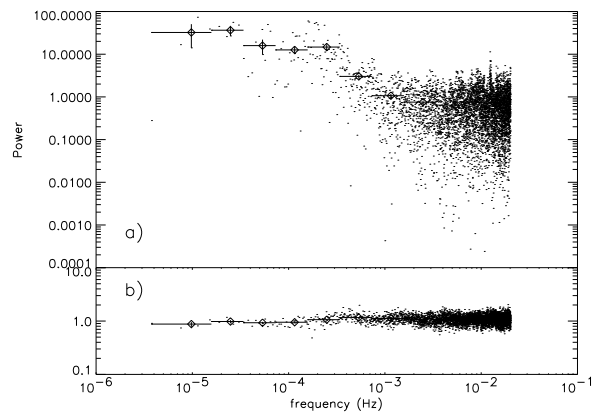,width=8cm}
\caption[ ]{a) Periodogram of the ROSAT PSPC observation of October 1992
(see Fig.3), b) Sample periodogram of 100 simulated white noise lightcurves
with the same sampling pattern as the original data.}
\end{figure}

The short term behavior is analyzed by computing the sample periodogram of
the lightcurves of the 16 individual orbits (Fig.5). At frequencies lower
than 0.002 Hz the power spectrum exhibits a weak linear trend which can be
interpreted as the begin of the red noise behavior. Since the background
countrate of the ROSAT PSPC detector is very low (Pfeffermann et al. 1986),
the flat power spectrum at high frequencies indicates the white noise 
behavior of the X-ray source NGC~6814 on timescales shorter than 500 sec
in the observation of October 1992.

We have also applied SSM fits to the continuous lightcurves of the two
longest individual orbits (Fig.6) to examine the short term
variability. The SSM is an alternative model to analyze the variability
seen in the X-ray lightcurves in the frequency domain instead of the time
domain.  Those models are based on the theory of autoregressive processes
(Scargle 1981, Honerkamp 1993), denoted as AR[p], which usually cannot be
observed directly since the observational noise (i.e. detectors, particle
background) overlays the process powering the AGN. An SSM-AR[p] fit applied
to the time series data yields the dynamical parameters of the underlying
stochastic process. The number of terms $p$ used for the regression of the
time series determines the order of the AR process. Depending on the order,
the derived dynamical parameters represent damped oscillators, pure
relaxators or its superpositions.

The NGC~6814 lightcurve of one orbit (Fig.6a) can be described by an
autoregressive process with a relaxator timescale of about 140 sec. This
timescale corresponds to the lifetime of the exponentially decaying
autocorrelation function. The other orbit (Fig.6b) is very likely a white
noise lightcurve. The Kolmogorov-Smirnov test prefers an SSM-AR[0] model
and the relaxation times of higher order AR fits equal the bin time which
indicates that these relaxators are negligible. Even though the statistical
significance of the SSM result is limited it is interesting to note that
they are in agreement with the variability timescale derived by applying
the Fourier analysis.

Due to the flux contamination of the close X-ray binary RXJ1940.1-1025 the
much longer EXOSAT ME or the Ginga data with a better duty cycle of NGC~6814
cannot be used for a temporal analysis of the galaxy's variability.

\begin{figure}
%\picplace{8 cm}
\centering
\psfig{file=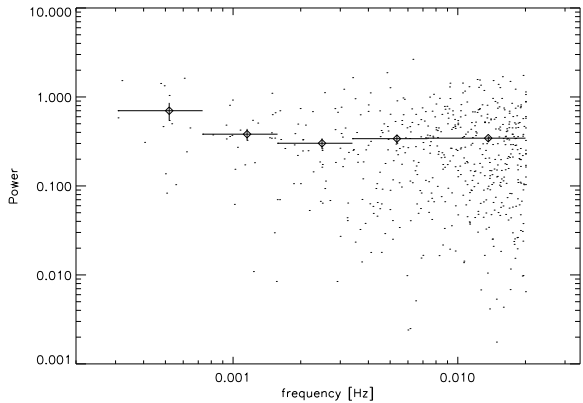,width=8cm}
\caption[ ]{Sample periodogram of the ROSAT PSPC observation of October
1992 (see Fig.3, compilation of the 16 individual orbit periodograms).}
\end{figure}

\begin{figure}
%\picplace{8 cm}
\centering
\psfig{file=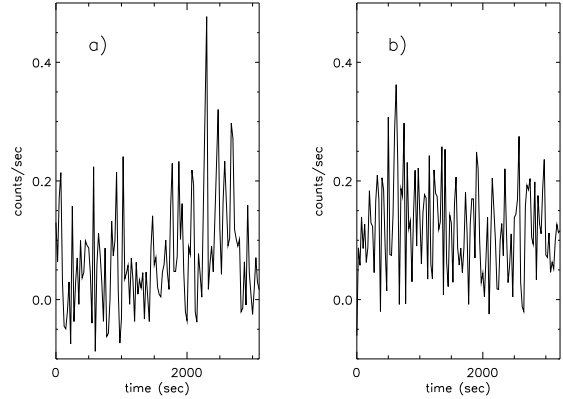,width=8cm}
\caption[ ]{Orbit lightcurves (orbit 1/16 and 3/16, 25 sec bins).}
\end{figure}

\begin{table}\centering
\caption{SSM Fit to orbit lightcurve (Fig.6a)}
\begin{tabular}[c]{ccccc}
\hline Model & R$^a$ & P$^b$ & ${\tau}^c$ & KS-test$^d$ \\ SSM AP[p] & &
(sec) & (sec) & \\ \hline 0 & 1 & - & - & 74.2\% \\ 1 & 0.859 & 0 & 137.4 &
96.7\% \\ 2 & 0.822 & 0 & 136.2 & 95.0\% \\ & & 0 & 5.1 & \\ \hline
\multicolumn{5}{l}{\small $^a$ Variance of the observational noise}\\
\multicolumn{5}{l}{\small $^b$ Damped oscillator and $^c$ relaxator
timescales}\\ \multicolumn{5 }{l}{\small $^d$ Kolmogorov-Smirnov test on
white noise residuals}\\
\end{tabular}
\end{table}

\begin{table}\centering
\caption{SSM fit to orbit lightcurve (Fig.6b)}
\begin{tabular}[c]{ccccc}
\hline Model & R$^a$ & P$^b$ & ${\tau}^c$ & KS-test$^d$ \\ & SSM AR[p]&
(sec) & (sec) & \\ \hline \\ 0 & 1 & - & - & 95.4\% \\ 1 & 0.857 & 0 & 25.1
& 77.4\% \\ 2 & 0.814 & 154.4 & 21.5 & 99.3\% \\ \hline
\multicolumn{5}{l}{\small $^a$ Variance of the observational noise}\\
\multicolumn{5}{l}{\small $^b$ Damped oscillator and $^c$ relaxator
timescales}\\ \multicolumn{5}{l}{\small $^d$ Kolmogorov-Smirnov test for
white noise residuals}\\
\end{tabular}
\end{table}

\section {Discussion} 

Despite the fact that the Seyfert galaxy NGC~6814 has lost its periodicity
it still remains one of the the most variable AGN with short term
variability on timescales of a few hundred seconds. If a homogeneous X-ray
emission region is assumed, this variability timescale is of the same order
as the light travel time across the innermost region of the AGN accretion
disk. The characterization of AGN variability by this single timescale is
misleading as this timescale strongly depends on the observed realization
of the lightcurve and therefore varies from one observation to the
next. Furthermore the AGN periodograms indicate that the variability is not
dominated by a single variability timescale but by a distribution of
timescales. The goal is to describe this distribution of all occuring
timescales in a more general way (McHardy 1988).

An alternative approach to the short term variability is to interpret the
observed X-ray variability as the superposition of single X-ray shots
generated in the emission process. This scenario starts with UV photons
which are generated by the inflow of accretion inhomogeneities. These UV
photon peaks trigger X-ray flares with a specific pulse profile by thermal
Comptonisation (Payne 1980, Liang and Nolan 1983). The relaxation time of
the exponentially decaying shots corresponds to the obtained relaxator
timescale. This relaxation time determines the distribution of frequencies
in the periodogram with the typical `red noise' behavior and the flattening
at low frequencies.


\begin{thebibliography} {999}
%
\bibitem [1978] {elvis}
Elvis M., Maccacaro T., Wilson A.S. et al., 1978, MNRAS 183, 129
\bibitem [1989] {elvisb}
Elvis M., Lockman F.J., Wilkes B.J., 1989, AJ 97, 777
\bibitem [1996] {fried}
Friedrich S., Staubert R., Lamer G. et al., 1996, A\&A 306, 860
\bibitem [1986] {pfeffer}
Pfeffermann E., et al., 1986, Proc SPIE 733, 519
\bibitem [1992]{green}
Hamilton J.D., 1995, Time Series Analysis, Princeton University Press
\bibitem [1987]{honerkamp}
Honerkamp J., 1993, Stochastic Dynamical Systems, VCH Publ. New York, Weinheim 
\bibitem [1987]{lawrence}
Lawrence A., Watson M.G., Pounds K.A. et al., 1987, Nature 325, 694
\bibitem [1989]{liang}
Liang E.P., Nolan P.L., 1983, Space Sci. Review 38, 353
\bibitem [1989]{madej}
Madejski G.M., Done C., Turner T.J., et al. 1993, Nature 365, 626
\bibitem [1987]{mchardy}
McHardy I., Czerny B., 1987, Nature 325, 696
\bibitem [1987]{mchardy88}
McHardy I., 1988, Mem.S.A.It. 59, 239
\bibitem [1996] {koenig}
K\"onig M., Timmer J., 1996, A\&A, accepted
\bibitem [1993]{koen}
Koen C., Lombard F., 1993, MNRAS 263, 287
\bibitem [1989]{payne}
Payne, D.G. 1989, ApJ, 237, 951
\bibitem [1994]{papadakis}
Papadakis I.E., Lawrence A., 1995, MNRAS 272, 161
\bibitem [1994]{patt}
Patterson J., Skillman D.R., Thorstensen J. et al., 1995, PASP 107, 307
\bibitem [1989]{priestley}
Robinson E.L., Nather R.E., 1979, ApJ Suppl. 39, 461
\bibitem [1996]{roth}
Rother F., 1996, private communication
\bibitem [1984]{scargle:81}
Scargle J.D., 1981, ApJ Suppl. 45, 1
\bibitem [1984]{staub}
Staubert R., K\"onig M., Friedrich. S. et al., 1994, A\&A 288, 513
\bibitem [1992]{turn}
Turner T.J., Done C., Mushotzky R.F., Madejski G., Kunieda H., 1992,
ApJ 391, 102
\bibitem [1994]{wat}
Watson M.G., Rosen S.R., O'Donoghue D. et al., 1995, MNRAS 273, 681
\bibitem [1989]{yaqoob}
Yaqoob T., Warwick R.S., Pounds K.A., 1989, MNRAS 236, 153
%
\end {thebibliography}
\end{document}